\documentclass[preprint,12pt,authoryear]{elsarticle}

\usepackage{amsthm} 
\usepackage{amssymb} 
\usepackage{amstext} 
\usepackage{amsmath} 
\usepackage{bm} 
\usepackage{graphicx} 
\usepackage{epstopdf} 
\usepackage{subcaption} 
\usepackage{float}
\usepackage{tikz}
\usepackage{lineno,hyperref}
\usepackage[capitalize]{cleveref} 
\usepackage[makeroom]{cancel} 
\usepackage{multirow}

\modulolinenumbers[5]

\crefalias{subequation}{equation} 
\crefalias{eqnarray}{equation} 
\crefformat{pluraleq}{Eqs.~(#2#1#3)} 

\journal{arXiv}

\biboptions{authoryear}

\begin{document}

\begin{frontmatter}

\title{A nonstiff solution for the stochastic neutron point kinetics equations}

\author[ufrgs]{M. Wollmann da Silva\fnref{milena}}
\author[ucb]{R. Vasques\corref{cor1}}
\author[ufrgs]{B.E.J. Bodmann\fnref{bodmann}}
\author[ufrgs]{M.T. Vilhena\fnref{vilhena}}

\address[ufrgs]{UFRGS - Federal University of Rio Grande do Sul, Av. Osvaldo
Aranha 99, 90046-900\\ Porto Alegre, RS, Brazil}
\address[ucb]{University of California, Berkeley, Department of Nuclear Engineering, 4155 Etcheverry Hall \\ Berkeley, CA 94720-1730}

\cortext[cor1]{Corresponding author: richard.vasques@fulbrightmail.org\\
Postal address: University of California, Berkeley, Department of Nuclear Engineering, 4103 Etcheverry Hall, Berkeley, CA 94720-1730}
\fntext[milena]{milena.wollmann@ufrgs.br}
\fntext[bodmann]{bardobodmann@ufrgs.br}
\fntext[vilhena]{vilhena@mat.ufrgs.br}

\begin{abstract}

We propose an approach to solve the stochastic neutron point kinetics equations using an adaptation of the diagonalization-decomposition method (DDM).
This new approach (Double-DDM) yields a nonstiff solution for the stochastic formulation, allowing the calculation of the neutron and precursor densities at any time of interest without the need of using progressive time steps.
We use Double-DDM to compute results for stochastic problems with constant, linear, and sinusoidal reactivities.
We show that these results strongly agree with those obtained by other approaches established in the literature.
We also compute and analyze the first four statistical moments of the solutions.

\end{abstract}

\begin{keyword}
point kinetics \sep stochastic moments \sep stiffness \sep decomposition method
\end{keyword}

\end{frontmatter}

\section{Introduction}\label{sec1}
\setcounter{section}{1}
\setcounter{equation}{0} 

The neutron point kinetics equations \citep{hetrick_71, kinard_04, hayes_05} are the coupled differential equations for the neutron density and for the delayed neutron precursor concentrations. 
These equations model the time-dependent behavior of a nuclear reactor and  provide insight into the dynamics of its operation.
The time-dependent parameters in this system are the reactivity function and the neutron source term.

The neutron density and delayed neutron precursor concentrations vary randomly with time; however, the point kinetics equations are deterministic and can only be used to estimate average values.
Random fluctuations in the neutron density and precursor concentrations can be significant at low power levels \citep{hurwitz_63a,hurwitz_63b}, which points to the importance of estimating these variations.

\citet{hayes_05} have generalized the standard deterministic point kinetics equations, deriving a system of stochastic differential equations that model the random behavior of the neutron density and the precursor concentrations in a point reactor.
Due to the issue of stiffness, this system was implemented numerically using a stochastic piecewise constant approximation method (Stochastic PCA).
Work performed by \citet{saha_12} has shown that order 1.5 strong Taylor and Euler-Maruyama numerical methods are valid computational alternatives to Stochastic PCA in solving the stochastic point kinetics equations.
However, with the exception of cases modeled with either none or only one group of neutron precursors, the stiffness of the problem remains.

In this paper we propose to solve this stochastic formulation using a double decomposition approach based on the diagonalization-decomposition method (DDM) decribed by \citet{wollmann_14}.
This proposed method is the major novelty and principal contribution of this work, yielding a {\em nonstiff} solution for the stochastic point kinetics equations.
Specifically, this approach allows the calculation of the neutron and precursor densities at any time of interest without the need of using progressive time steps.
This solution is obtained with a minimal amount of numerical approximations of the model; the largest numerical effort lies in the truncation of the decomposition and  the integrations required by DDM.

The major caveat in this approach is that convergence of DDM is yet to be proven.
For this reason, a Lyapunov criterion \citep{boichenko_05} is used to guarantee convergence (cf. \citeauthor{petersen_11a}, \citeyear{petersen_11a}; \citeauthor{wollmann_14}, \citeyear{wollmann_14}).
We present computational results for problems with constant, linear, and sinusoidal reactivities. 
The results of the proposed method are compared against those of other approaches established in the literature, showing strong agreement.
We also compute the first four statistical moments of the solutions.

This work is an expanded version of a recent conference paper \citep{wollmann_15b}.
The remainder of this paper is organized as follows.
In \cref{sec2} we present a brief review on the key aspects of DDM.
In \cref{sec3} we formulate the stochastic point kinetics equations.
We introduce the proposed double decomposition approach in \cref{sec4}.
Numerical results are given in \cref{sec5} for problems with constant (\cref{sec51}) and time-dependent (\cref{sec52}) reactivities.
The paper concludes in \cref{sec6} with a discussion of the work presented.

\section{The Diagonalization-Decomposition Method (DDM)}
\label{sec2}
\setcounter{section}{2}
\setcounter{equation}{0}

Following \citet{wollmann_14}, one can obtain an analytical representation for the solution of the neutron point kinetics equations that is free of stiffness.
The neutron point kinetics equations with six groups of precursors and time-dependent reactivity $\rho(t)$ are written as:
\begin{subequations}\label[pluraleq]{eq2.1}
\begin{align}
\frac{d}{dt}n(t)&=\frac{\rho(t)-\beta}{\Lambda} n(t)
+\sum_{i=1}^{6}\lambda_{i}C_{i}(t),  \qquad n(0)=n_{0} \;, 
\\
\frac{d}{dt}C_{i}(t)&=\frac{\beta_{i}}{\Lambda} n(t)-\lambda_{i}C_{i}(t),\hspace{72pt}
C_{i}(0)=\frac{\beta_{i}n_{0}}{\Lambda\lambda_{i}}\;,
\end{align}
\end{subequations}
for $i=1,2,...,6$.
Here, $n(t)$ is the neutron density; $C_{i}(t)$ is the density of the i$^{th}$ delayed neutron precursor group; $\lambda_i$ is the decay constant for a specific group $i$; $\Lambda$ represents the neutron mean generation time; and $\beta_i$ represents the delayed-neutron fraction in a specific group $i$.
The total fraction of delayed neutrons is given by $\beta = \sum\limits_{i=1}^6\beta_i$.

A recursive scheme with finite recursive depth $R$ is used to obtain a solution. The truncation index is determined with exponential convergence by the Lyapunov criterion \citep{boichenko_05, petersen_11a}, evaluated after each recursion step.
The neutron population and the precursors concentration are written in terms of the solution from a recursion initialization ($j=0$) and the respective correction terms ($j>0$) for an appropriate $R \in \mathbb{N}$: 
\begin{subequations}\label[pluraleq]{eq2.2}
\begin{align}
n(t)&=\sum_{j=0}^{R}n_{j}(t)\,,\\
C_{i}(t)&=\sum_{j=0}^{R}C_{i,j}(t)\, .
\end{align}
\end{subequations}

The combination of \cref{eq2.1} and (\labelcref{eq2.2}) yields a system with $7\times R$ unknowns.
We define
\begin{subequations}\label[pluraleq]{eq2.3}
\begin{align}
\bm Y(t) &= \sum_{j=1}^R \bm Y_j(t)\,,\\
\bm Y_j(t) &=[n_j(t),C_{1,j}(t),C_{2,j}(t),C_{3,j}(t),C_{4,j}(t),C_{5,
j}(t),C_{6,j}(t)]^T\,,\\
\bm\Omega &=\text{diag}\left(\frac{\rho_{0}-\beta}{\Lambda},-\lambda_1, -\lambda_2, -\lambda_3, -\lambda_4, -\lambda_5, -\lambda_6 \right)\,,\label{eq2.3c}
\end{align}
and
\begin{align}
\bm\Xi &= \left[
\begin{array}{cc} \vspace{0.25cm}
\displaystyle{\frac{\rho_{1}(t)}{\Lambda}}&\{\lambda_i\} \\
\vspace{0.25cm}
 \displaystyle{\left\{\frac{\beta_i}{\Lambda}\right\}^T} & \bm 0\\
 \end{array}
 \right] \, , \label{eq2.3d}
\end{align}
\end{subequations}
where $\rho_0$ (constant) and $\rho_1(t)$ are such that $\rho(t) = \rho_0 + \rho_1(t)$. Given the recursive system
\begin{subequations}\label[pluraleq]{eq2.4}
\begin{align}
& \frac{d}{dt}\bm Y_0(t)-\bm\Omega \bm Y_0(t)=0\,,\label{eq2.4a}\\ 
& \frac{d}{dt}\bm Y_j (t)-\bm\Omega \bm Y_j (t)=
\bm\Xi(t)\bm Y_{j-1}(t) \;, \qquad j>0\;\label{eq2.4b},
\end{align}
\end{subequations}
the solution of \cref{eq2.4a} is
\begin{subequations}\label[pluraleq]{eq2.5}
\begin{align}\label{eq2.5a}
\bm Y_0(t) = \exp(\bm\Omega t)\bm Y_0(0)\,,
\end{align}
with $\bm Y_0(0) = [n_0, C_1(0),C_2(0),C_3(0),C_4(0),C_5(0),C_6(0)]^T$.
\Cref{eq2.4b} may be formally solved by the Laplace transform:
\begin{align}\label{eq2.5b}
 \bm Y_j(t)&=\exp(\bm\Omega t)\cancelto{0}{\bm Y_j(0)}+\int_{0}^{t}\exp(\bm\Omega
\tau)\bm\Xi(t-\tau)\bm Y_{j-1}(t-\tau)d\tau \;, \qquad j>0\;,
\end{align}
\end{subequations}
since the initial condition from \cref{eq2.1} is fully absorbed in \cref{eq2.5a}. 
The integral in \cref{eq2.5b} is evaluated using the Gauss-Legendre method.

\begin{figure}[H]
 \centering
 \includegraphics[scale=1]{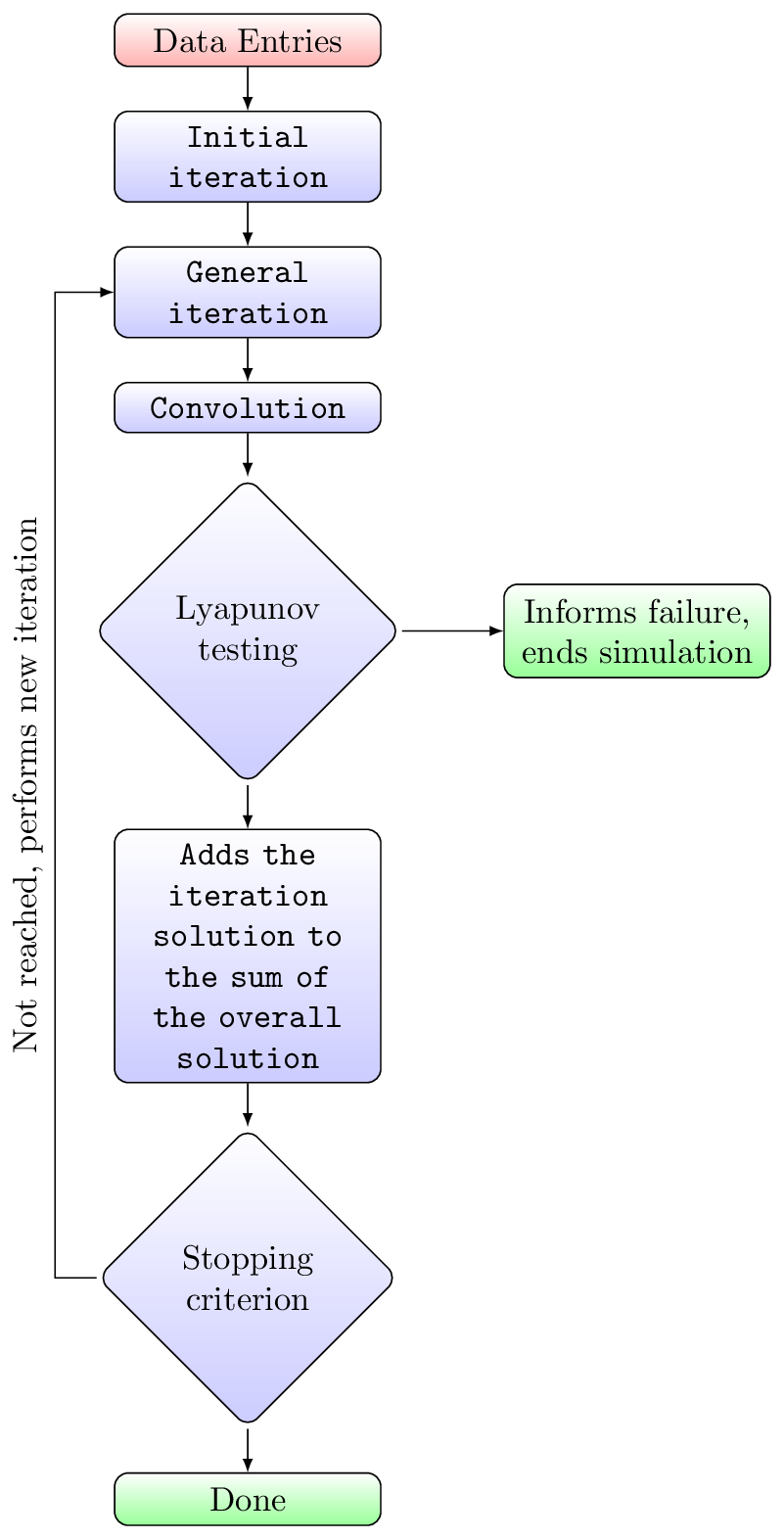} 
\caption{\footnotesize{DDM approach to solve the deterministic problem}}\label{fig1}
\end{figure}

A flowchart describing the implementation of this method is given in \cref{fig1}.
The solution is obtained in an analytical representation that may be evaluated for any time value (free of stiffness).

\section{The Stochastic Formulation}
\label{sec3}
\setcounter{section}{3}
\setcounter{equation}{0}

\citet{hayes_05} derived a system of It\^o stochastic differential equations that model the dynamics of the neutron density and the delayed neutron precursors in a point nuclear reactor.
This formulation describes the variation of the population and can be interpreted as a balance between deaths, births, and transformations of neutrons in the system.
The probabilities of these events are determined by the physical parameters of the model, such as the total and partial delayed neutron fraction; the fraction of delayed neutrons of each precursor group; the decay constant of each group; and the average number of neutrons produced in each fission.

Assuming a time interval small enough such that only one event occurs, one can write
\begin{subequations}\label[pluraleq]{eq3.1}
\begin{align}\label{eq3.1a}
 \frac{d}{dt}\bm Y(t)=\bm{\hat{A}}(t) \bm Y(t)+
 \bm Q(t)
 +\bm{\hat{B}}^\frac{1}{2}(t)\frac{d}{dt}\bm{W}(t) \;, 
\end{align}
where 
\begin{align}
\bm Y(t) &=[n(t),C_{1}(t),C_{2}(t),C_{3}(t),C_{4}(t),C_{5}(t),C_{6}(t)]^T\,,\\
\bm Q(t) &=[q(t),0,0,0,0,0,0]^T\,,\\
\bm W(t) &=[W_0(t),W_1(t),W_2(t),W_3(t),W_4(t),W_5(t),W_6(t)]^T\,,\\
\bm{\hat{A}}(t)&=\bm\Omega + \bm\Xi(t)\,, \vspace{5pt}\\
\bm{\hat{B}}(t)&=\left[\begin{array}{ccccc}
 \zeta & a_1& a_2&\dots &a_6\\
 a_1 & b_{1,1} & b_{1,2} &\dots & b_{1,6}\\
 a_2 & b_{2,1} &b_{2,2} &\dots & b_{2,6}\\
 \vdots & \vdots & \ddots & \ddots & \vdots\\
 a_6&b_{6,1}&\dots&b_{6,5}&b_{6,6}
 \end{array}\right]\, .
\end{align}
Here, $\bm\Omega$ and $\bm\Xi(t)$ are given by \cref{eq2.3c,eq2.3d}, $W_i(t)$ are Wiener processes, and
\begin{align}
\zeta &=\left(\frac{2\beta-1-\rho(t)+(1-\beta)^2\nu}{\Lambda}\right) n(t) \sum_{i=1}^{6}\lambda_i C_i(t)+q(t)\;,\\
 a_i &=\frac{\beta_i}{\Lambda}\big(-1+(1-\beta)\nu\big)n(t)-\lambda_iC_i(t) \;,\\
 b_{i,j}&=\frac{\beta_i\beta_j\,\nu}{\Lambda}n(t)+\delta_{ij}\lambda_iC_i(t) \; ,
\end{align}
\end{subequations}
with $\nu =$ number of neutrons per fission.   
Note that if $\bm{\hat{B}}(t)$ = 0, then \cref{eq3.1a} (with $\bm Q(t)=0$) reduces to the deterministic problem discussed in \cref{sec2}.

\section{The Proposed Method (Double-DDM)}
\label{sec4}
\setcounter{section}{4}
\setcounter{equation}{0} 

We propose to solve the stochastic formulation in \cref{sec3} by adapting the recursive method described in \cref{sec2}.
This approach yields a nonstiff solution to the stochastic system in \cref{eq3.1}.

Since the matrix $\bm{\hat{B}}$ depends on both the neutron populations
and the delayed neutron concentrations, we resort to a double decomposition to obtain a solution for this problem:
\begin{itemize}
\item[I.] Using DDM as described in \cref{sec2}, \cref{eq3.1a} is solved for $\bm{\hat{B}}=0$; this yields the deterministic solution $\bm Y(t)=\bm Y_d(t)$;
\item[II.] $\bm Y_d(t)$ is used to determine the matrix $\bm{\hat{B}}$ for a sequence of discrete time steps (its components are constant in each time step);
\item[III.] $\bm{\hat{B}}^\frac{1}{2}$ is obtained through diagonalization ($\bm{\hat{B}}$ is symmetric);
\item[IV.] Since $\bm Q(t)$ is known for a specific time interval, a decomposition scheme analogue to DDM is applied:
\begin{subequations}\label[pluraleq]{eq4.1}
\begin{align}
 \frac{d}{dt}\bm Y_0(t)-\bm\Omega \bm Y_0(t)&=0\,,\\ 
\frac{d}{dt}\bm Y_j (t)-\bm\Omega \bm Y_j (t)&=
\bm{\Xi}(t)\bm Y_{j-1}(t)+\bm{{\cal{F}}}(t) \; ,
\end{align}
\end{subequations}
where $\bm{{\cal{F}}}(t) =\bm Q(t)+\bm{\hat{B}}^\frac{1}{2}\frac{d}{dt} {\bm{W}}(t)$ are constants known in each time step.
\end{itemize}

\begin{figure}[H]
\centering
 \includegraphics[scale=1]{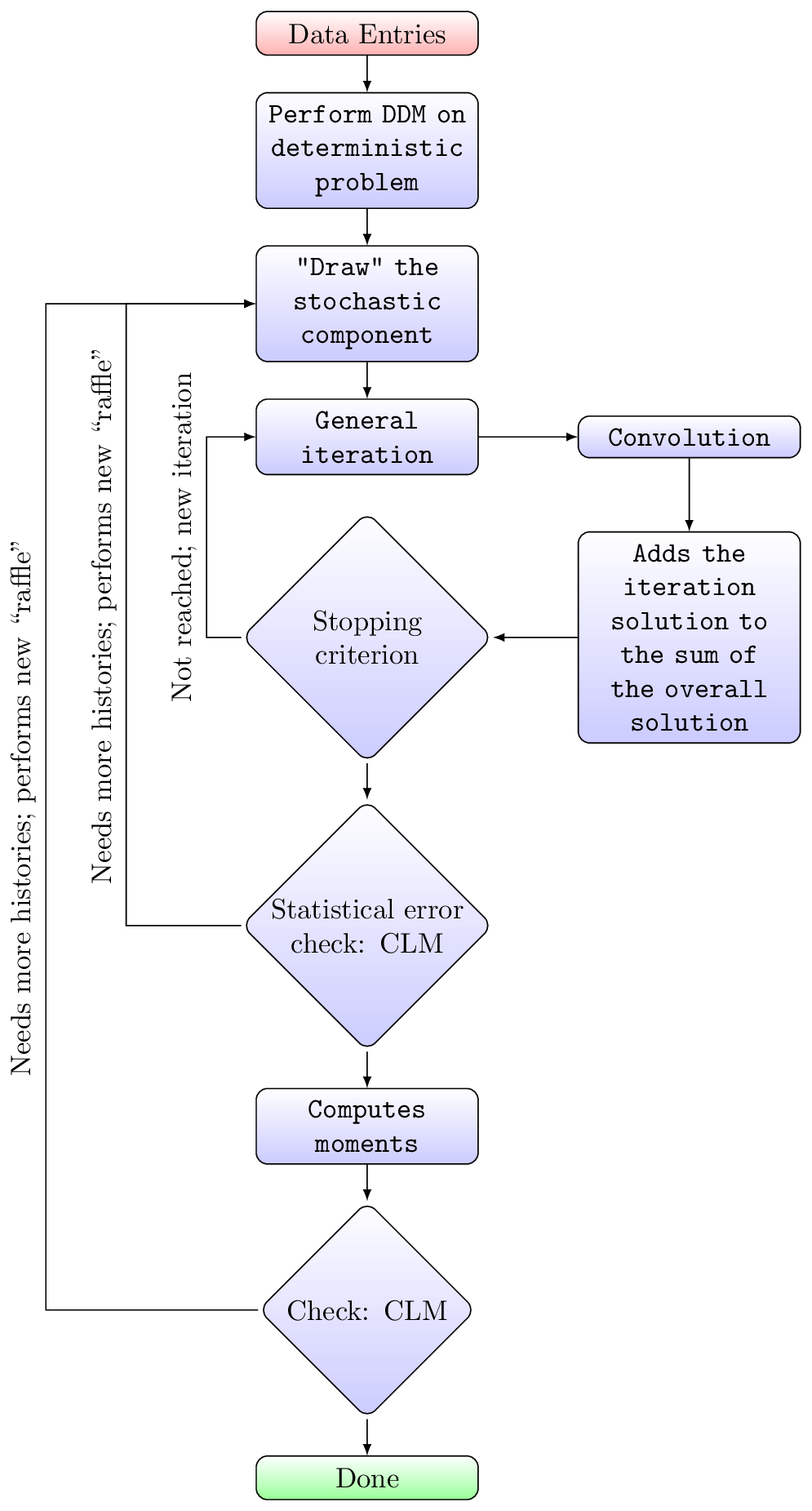} 
\caption{\footnotesize{Double-DDM approach to solve the stochastic problem}}\label{fig2}
\end{figure}

The total number $K$ of stochastic components ``drawn" in this approach is defined by the Central Limit Theorem (CLM) to guarantee a small statistical error for the first four moments.
For all results shown in this paper, $K$ is large enough to guarantee a statistical error smaller than 0.05\% with 95\% confidence. 
The flowchart in \cref{fig2} describes the implementation of the proposed method.

\section{Numerical Results}
\label{sec5}
\setcounter{section}{5}
\setcounter{equation}{0} 

In this section we present numerical results for the Double-DDM approach proposed in \cref{sec4} for examples with (i) constant and (ii) time-dependent reactivities.
We compare these results against those obtained with other approaches established in the literature.
We calculate the expected value $E(\bm Y)$ and variance Var$(\bm Y$), given by
\begin{subequations}
\begin{align}
E(\bm Y) &= \sum_{k=1}^K \frac{\bm Y^{(k)}}{K}\,,\\
\text{Var}(\bm Y) = \sigma^2(\bm Y) &= \sum_{k=1}^K \frac{\left(\bm Y^{(k)}-E(\bm Y)\right)^2}{K}\,,
\end{align}
where the index $k$ represents the different choices of stochastic components (histories).
We also calculate skewness and excess kurtosis for the neutron density $n$, defined as
\begin{align}
\text{Skew}(n) &= \gamma_1(n) =\sum_{k=1}^{K}\frac{\left(n^{(k)}- E(n)\right)^3}{K\sigma^3(n)}\,,\\
 \text{Kurt}(n)-3&=\gamma_2(n) = \sum_{k=1}^{K}\left[\frac{\left(n^{(k)}- E(n)\right)^4}{K\sigma^4(n)}\right]-3\,,
\end{align}
\end{subequations}
which gives us further insight on the behavior of the stochastic solutions.

\subsection{Constant Reactivity}\label{sec51}

In the following examples we present the results of four methods established in the literature: Monte Carlo and Stochastic PCA \citep{hayes_05}; order 1.5 strong Taylor and Euler-Maruyama \citep{saha_12}.
These results are reproduced here as they were reported in the aforementioned references.

We compare these results with those obtained with the deterministic diagonalization-decomposition method (DDM) and with the Double-DDM approach.
The solutions of the deterministic DDM are obtained by solving \cref{eq3.1a} with $\bm{\hat{B}}=0$; we point out that this is also the first step of Double-DDM.

\subsubsection{Step-reactivity insertion: one precursor}\label{sec511}

This example does not model an actual physical nuclear reactor problem.
Nevertheless, considering only one group of precursors implies that the stiffness of the problem disappears; this provides a simple computational solution that is useful for a first comparison with other methods.

The physical parameters are $\lambda_1=0.1$,  $\beta=\beta_1=0.05$, $\nu=2.5$, $q=200$, $\Lambda=\frac{2}{3}$, and $\rho(t)=-\frac{1}{3}$ for $t\geq 0$.
The initial condition is $\bm{Y}(0)=[400, 300]^T$.
The expected values and standard deviations of $n(t)$ and $C_1(t)$ at time $t=2$ are presented in \cref{tab1} for each of the methods.
\begin{table}[H]
\caption{Results for one group of precursors and step-reactivity insertion $\rho(t)=-1/3$.}\label{tab1}
\centering{\small{
\begin{tabular}{ccccccc} \hline\hline
 									& {\bf Monte} 	& {\bf Stochastic}	 & {\bf Euler}         & {\bf Taylor}		&  \multirow{2}{*}{{\bf DDM}}		&{\bf Double} \\ 
 									& {\bf Carlo}  	& {\bf PCA}			 & {\bf Maruyama}  & {\bf 1.5} 		&  	  		& {\bf DDM}\\ \hline
 ${E(n(2))}$ 				&400.03 	& 395.32     &412.23		  &412.10		&412.13  &402.35\\
${\sigma(n(2))}$ 		&27.311 	& 29.411     &34.391		  &34.519	& --			&28.610\\ 
${E(C_1(2))}$ 			& 300.00 & 300.67     &315.96		  &315.93	&315.93	&305.84\\
${\sigma(C_1 (2))}$ 	& 7.8073 & 8.3564     &8.2656 	  &8.3158	& --			&7.9240 \\ \hline\hline
\end{tabular}
}}
\end{table}
A total of $K=1,000$ histories were accumulated for the Double-DDM approach.
Skewness and excess kurtosis for the neutron density were found to be $\gamma_1(n(2)) = -1\times 10^{-10}$ and $\gamma_2(n(2)) = 3\times 10^{-11}$.
The fact that $\gamma_1$ and $\gamma_2$ are nearly zero implies that the distribution of stochastic solutions is symmetric and has Gaussian-like peak and tail.

It can be seen that there exists a close agreement between Double-DDM and the results obtained with Monte Carlo and Stochastic PCA.
Euler-Maruyama and order 1.5 strong Taylor (Taylor 1.5) yield slightly higher results, very close to those obtained with the deterministic DDM.

\subsubsection{Step-reactivity insertion: six precursors}\label{sec512}

The following two examples model step-reactivity insertions for a thermal nuclear reactor.
In this case, we consider a stiff system of equations with six precursor groups.
The set of physical parameters is taken from \citet{kinard_04}:
$\nu=2.5$, $\Lambda=0.00002$, $q=0$, and $\beta=7\times 10^{-3}$, with $\beta_i$ and $\lambda_i$ given in \cref{tab2}. The initial condition is given by
\begin{align}
\bm{Y}(0)=100 \left[1, \frac{\beta_1}{\lambda_1 \Lambda}, \frac{\beta_2}{\lambda_2 \Lambda}, \frac{\beta_3}{\lambda_3 \Lambda}, \frac{\beta_4}{\lambda_4 \Lambda}, \frac{\beta_5}{\lambda_5 \Lambda}, \frac{\beta_6}{\lambda_6 \Lambda}\right]^T\,.
\end{align}
\begin{table}[H]
\caption{Fraction of delayed neutrons and decay constants for the precursor groups.}\label{tab2}
\centering{\small{
\begin{tabular}{lllllll}
{\bf Group}							& {\bf 1}	&{\bf 2}	&{\bf 3}	&{\bf 4}	&{\bf 5}	&{\bf 6}\\\hline
$\beta_i\times 10^{-3}$	&0.266	&1.491	&1.316	&2.849	&0.896	&0.182 \\ \hline
$\lambda_i$						&0.0127	&0.0317	&0.115	&0.311	&1.4			&3.87 \\
\hline
\end{tabular}
}}
\end{table}

We compute results for a prompt subcritical step-reactivity insertion $\rho(t)=0.003$ at time $t=0.1$, and for a prompt critical step-reactivity insertion $\rho(t)=0.007$ at time $t=0.001$.
We define
\begin{align}
C(t)=\sum_{i=1}^6 C_i(t)\,,
\end{align}
and present the expected values and standard deviations for each of the methods in \cref{tab3}.
\begin{table}[H]
\caption{Results for six groups of precursors with subcritical $(\rho(t)=0.003)$ and critical $(\rho(t)=0.007)$ step-reactivity insertions.}\label{tab3}
\centering{\small{
\begin{tabular}{cccccccc} \hline\hline
\multirow{2}{*}{{$\bm\rho$}}& 									& {\bf Monte} 	& {\bf Stochastic}	 & {\bf Euler}         & {\bf Taylor}		&  \multirow{2}{*}{{\bf DDM}}		&{\bf Double} \\ 
& 									& {\bf Carlo}  	& {\bf PCA}			 & {\bf Maruyama}  & {\bf 1.5} 		&  	  		& {\bf DDM}\\ \hline
\multirow{4}{*}{$0.003$}
&$E(n(0.1))$ 						&183.04 		& 186.31     &208.60		&199.41	&200.01  	&187.05\\
&$\sigma(n(0.1))$ 				&168.79 		& 164.16     &255.95		&168.55	& --				&167.83\\ 
&$E(C(0.1))\times 10^5$ 	& 4.478 		& 4.491  		&4.498		&4.497		&4.497		&4.488\\
&${\sigma(C(0.1))}$ 			& 1495.7 	& 1917.2    	&1233.4	 	&1218.8		& --				&1475.6 \\ \hline\hline
\multirow{4}{*}{$0.007$}
&$E(n(0.001))$ 						&135.67 		& 134.55     &139.57		&139.57	&139.61  	&135.86\\
&$\sigma(n(0.001))$ 				&93.376 		& 91.242     &92.042		&92.047	& --				&93.210\\ 
&$E(C(0.001))\times 10^5$ 	& 4.464 		& 4.464  		&4.463		&4.463		&4.463		&4.463\\
&${\sigma(C(0.001))}$ 			& 16.226 	& 19.444    	&6.071	 	&18.337		& --				&17.845 \\ \hline\hline
\end{tabular}
}
}\end{table}
We collected $K=1,000$ histories for the subcritical step-reactivity $\rho(t)=0.003$.
Skewness was found to be $\gamma_1(n(0.1)) = -1\times 10^{-7}$, and excess kurtosis $\gamma_2(n(0.1)) = 1.3\times 10^{-9}$.
For the critical step-reactivity $\rho(t)=0.007$, $K=5,000$ histories were collected.
Skewness and excess kurtosis were computed respectively as $\gamma_1(n(0.001))=-1.02\times 10^{-7}$ and $\gamma_2(n(0.001))=1.15\times 10^{-8}$.
These results for the third and fourth moments indicate that the distribution of the stochastic solutions is symmetric and has neither a sharp peak nor a heavy tail.

As in the previous example, the results obtained with Double-DDM are in close agreement to the results from Monte Carlo and Stochastic PCA. The results from Euler-Maruyama and order 1.5 strong Taylor are closer to those of deterministic DDM.

\subsection{Time-Dependent Reactivity}\label{sec52}

The current literature lacks numerical results for the stochastic system in \cref{eq3.1a} with time-dependent reactivities.
For this reason, the results collected from the literature and reproduced next represent only the deterministic solution (with $\bm{\hat{B}}=0$). 
Although not ideal, this approach allows us to verify that the expected value obtained with Double-DDM closely agrees with well established models for problems with time-dependent reactivities.
All the following examples take into account six precursor groups.

\subsubsection{Linear reactivity $\rho(t)=at$}\label{sec521}

The following two examples model a ramp reactivity $\rho(t)=at$ for a thermal nuclear reactor.
The physical parameters considered are: $\Lambda=0.00001$, $n(0)=1.0$, and $\beta=6.403\times 10^{-3}$, with $\beta_i$ and $\lambda_i$ taken from \citet{lewins_78} and given in \cref{tab4}.
\begin{table}[H]
\caption{Fraction of delayed neutrons and decay constants for the precursor groups.}\label{tab4}
\centering{\small{
\begin{tabular}{lllllll}
{\bf Group}							& {\bf 1}	&{\bf 2}	&{\bf 3}	&{\bf 4}	&{\bf 5}	&{\bf 6}\\\hline
$\beta_i\times 10^{-3}$	&0.246	&1.363	&1.203	&2.605	&0.819	&0.167 \\ \hline
$\lambda_i$						&0.0127	&0.0317	&0.115	&0.311	&1.4			&3.87 \\
\hline
\end{tabular}
}}
\end{table}
We compute the neutron density $n(t)$ for two different choices of constant $a$: $0.25$ and $0.5$.
The results obtained with Double-DDM are given in \cref{tab5} for times $t=0.25, 0.5, 0.75,$ and $1.0$.
We compare these results with those obtained with the Pad\'e approximation \citep{aboanber_02} and the generalization of the analytical exponential model (GAEM), as reported by \citet{nahla_08}.
\begin{table}[H]
\caption{Neutron density $n(t)$ with ramp reactivity $\rho(t)=at$.}\label{tab5}
\centering{\small{
\begin{tabular}{cccccc} \hline\hline
\multirow{2}{*}{{$\bm a$}}& \multirow{2}{*}{{\bf Time}} & \multirow{2}{*}{{\bf Pad\'e}} &\multirow{2}{*}{{\bf GAEM}} &	\multirow{2}{*}{{\bf DDM}} & {\bf Double} \\ 
 & & & & & {\bf DDM} \\ \hline\
\multirow{4}{*}{$0.25$}
&0.25 & 1.069840 & 1.069541 & 1.069542 & 1.069763\\
&0.50 & 1.157065 & 1.156694 & 1.156695 & 1.157867\\
&0.75 & 1.265795 & 1.265331 & 1.265332 & 1.269374\\
&1.0   & 1.402562 & 1.401981 & 1.401982 & 1.403561\\\hline\hline
\multirow{4}{*}{$0.5$}
&0.25 & 1.149544 & 1.149200 & 1.149210 & 1.137216\\
&0.50 & 1.369438 & 1.368927 & 1.368928 & 1.356934\\
&0.75 & 1.708411 & 1.707600 & 1.707601 & 1.695607\\
&1.0   & 2.276692 & 2.275271 & 2.275272 & 2.263278\\\hline\hline
\end{tabular}
}
}\end{table}
The number of histories collected for the case $a=0.25$ was $K=15,738$.
We computed the higher moments for time $t=1.0$, finding the standard deviation $\sigma(n(1))=0.978$, skewness $\gamma_1(n(1))=-3.02\times 10^{-7}$, and excess kurtosis $\gamma_2(n(1))=-1 \times 10^{-7}$.
For the case $a=0.5$, we collected $K=27,523$ histories, and found 
$\sigma(n(1))=1.11345$, $\gamma_1(n(1))=-2.15\times 10^{-6}$, $\gamma_2(n(1))=-3 \times 10^{-8}$.
This shows that, in both cases, the distribution of stochastic solutions is nearly normal.

Double-DDM shows good agreement with the other methods shown in \cref{tab5}.
The results obtained with Double-DDM are slightly larger for the first case ($a=0.25$), and slightly smaller for the second case.
The relative differences are around $1\%$ or smaller, being well within 1 standard deviation.

\subsubsection{Sinusoidal reactivity $\rho(t)=b\sin(t)$}\label{sec522}

The last example simulates a case with sinusoidal reactivity $\rho(t)=b\sin(t)$, with $b=0.00073$, $\Lambda = 0.00003$, and $n(0)=1.0$. The total fraction of delayed neutrons is given by $\beta=6.473\times 10^{-3}$, with $\beta_i$ and $\lambda_i$ shown in \cref{tab6}.
\begin{table}[H]
\caption{Fraction of delayed neutrons and decay constants for the precursor groups.}\label{tab6}
\centering{\small{
\begin{tabular}{lllllll}
{\bf Group}							& {\bf 1}	&{\bf 2}	&{\bf 3}	&{\bf 4}	&{\bf 5}	&{\bf 6}\\\hline
$\beta_i\times 10^{-3}$	&0.214	&1.423	&1.247	&2.568	&0.748	&0.273 \\ \hline
$\lambda_i$						&0.0124	&0.0305	&0.111	&0.301	&1.14			&3.01 \\
\hline
\end{tabular}
}}
\end{table}
The results obtained with Double-DDM are presented in \cref{tab7} for every whole second up until $t=10$.
We compare these results with those reported in \citep{wollmann_14}.
These were obtained with the method introduced by \citet{kang_73}, referred to as K \& H in \cref{tab7}, and with the method of Taylor series (cf. \citeauthor{nahla_11}, \citeyear{nahla_11}).
\begin{table}[H]
\caption{Neutron density $n(t)$ with sinusoidal reactivity $\rho(t)=0.00073\sin(t)$.}\label{tab7}
\centering{\small{
\begin{tabular}{ccccc} \hline\hline
\multirow{2}{*}{{\bf Time}} & \multirow{2}{*}{{\bf K \& H}} &\multirow{2}{*}{{\bf Taylor}} &	\multirow{2}{*}{{\bf DDM}} & {\bf Double} \\ 
& & & & {\bf DDM} \\ \hline
0   & 1.00000 & 1.00000 & 1.00000 & 1.0000000\\
1   & 1.12397 & 1.12394 & 1.12396 & 1.1119659\\
2   & 1.16881 & 1.16884 & 1.16889 & 1.1568959\\
3   & 1.07443 & 1.07442 & 1.07448 & 1.0624859\\
4   & 0.95381 & 0.95380 & 0.95382 & 0.9418259\\
5   & 0.90737 & 0.90737 & 0.90735 & 0.8953559\\
6   & 0.96151 & 0.96158 & 0.96153 & 0.9495359\\
7   & 1.08748 & 1.08749 & 1.08745 & 1.0754559\\
8   & 1.17168 & 1.17164 & 1.17167 & 1.1596759\\
9   & 1.11128 & 1.11124 & 1.11130 & 1.0993059\\
10 & 0.98464 & 0.98464 & 0.98468 & 0.9726859\\\hline\hline
\end{tabular}
}
}\end{table}
We collected $K=2,934,237$ histories to achieve the requirement imposed for the statistical error (less than 0.05\% with 95\% confidence).
The higher moments for time $t=10$ yield $\sigma(n(10))= 1.3242$, $\gamma_1(n(10))=-0.0048$, and $\gamma_2(n(10))=-0.013$. These results indicate (i) a very small asymmetry in the distribution of stochastic solutions, with a slightly larger left tail; and (ii) a very small yet noticeable  ``flatter" peak when compared to a normal. 

\begin{figure}[H] 
 \centering
 \includegraphics[scale=1.0]{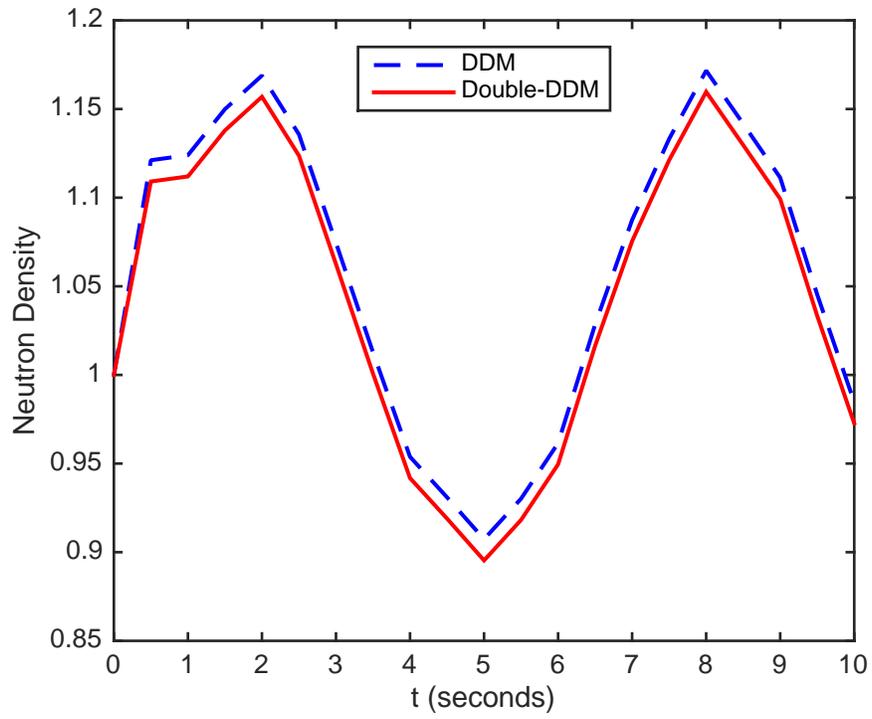} 
 \caption{\footnotesize{Neutron density for a sinusoidal reactivity $\rho(t) = 0.00073\sin(t)$}}\label{fig3}
\end{figure}

In general, Double-DDM closely agrees with the other methods presented here for comparison.
Results displayed in \cref{tab7} show that Double-DDM  yields slightly smaller results than those attained with the other methods.
This can be confirmed in \cref{fig3}, which depicts the average behavior of the stochastic solution compared to the deterministic solution (DDM).
These differences are, once again, very small ($\approx 1\%$), and well within 1 standard deviation.

\section{CONCLUSIONS}
\label{sec6}
\setcounter{section}{6} 
\setcounter{equation}{0} 

In this paper, we propose an approach to solve the stochastic neutron point kinetics equations with a solution procedure that is free of stiffness.
This is achieved through an adaptation of the diagonalization-decomposition method (DDM) introduced in \citet{wollmann_14}, wich provides a nonstiff solution for the deterministic point kinetics equations.
DDM uses the Laplace transform to obtain a formal solution, then applies a decomposition into a recursive scheme, using Gauss-Legendre integration.

The essential steps of the proposed approach (Double-DDM) are: (i) the deterministic problem is solved with DDM; (ii) the deterministic solution is used to build the stochastic component; (iii) another decomposition scheme analogue to DDM is used to solve the stochastic system.
This allows the calculation of the neutron density and precursor concentrations at any time of interest, without the need to resort to progressive time steps.
The elimination of stiffness comes from the fact that the evolution of the solution by recursion adds correction terms to the whole time interval of interest in each step, and simultaneously for each term that depends on a specific time scale.

Since convergence of DDM is yet to be proven, a Lyapunov criterion is used to guarantee convergence.
The results of the proposed method are compared against results obtained through other approaches established in the literature.
This comparison shows close agreement for problems with constant step-reactivity insertions, as well as time-dependent ramp reactivity and sinusoidal reactivity insertions.

In the current literature, the stochastic problem is mainly solved for constant reactivities; numerical solutions are limited to feasible time intervals due to the stiffness inherent to the problem.
The mitigation of the stiffness character in solving the stochastic formulation is the major novelty and principal contribution of this work.
Moreover, the analysis of the third and fourth moments of the stochastic solutions is, to the best of our knowledge, new.

The analyzed moments still depend on the size of the sample set and on the frequency with which the stochastic fluctuations are applied.
In principle, an adjustment such as variance reduction and its consequences for higher moments could yield more realistic results.
It would be necessary to find a reference scale in order to obtain such results independently of the sample size or frequency of application.
This still needs to be identified, as well as a necessary ingredient to mimic reactor fluctuations.
These tasks are left for future work.
 
\section*{Acknowledgments}
Milena Wollmann da Silva would like to thank the Coordination for the Improvement of Higher Education Personnel (CAPES, Brazil) for financial support.
Bardo E.J. Bodmann and Marco T\'ullio Vilhena thank the Federal University of Rio Grande do Sul (UFRGS) and also acknowledge financial support from the National Council of Scientific and Technological Development (CNPq, Brazil).
Richard Vasques prepared this paper under award number NRC-HQ-84-14-G-0052 from the Nuclear Regulatory Commission.
The statements, findings, conclusions, and recommendations are those of the authors and do not necessarily reflect the view of the U.S. Nuclear Regulatory Commission.

\section*{References}

\bibliography{refs}

\pagebreak

\noindent

\end{document}